\documentclass{llncs}
\usepackage{amssymb}
\usepackage{amsmath}
\usepackage{mathrsfs}
\usepackage{stmaryrd}
\usepackage{times}
\usepackage{latexsym}
\usepackage{hhline}
\usepackage{enumerate} 
\usepackage[latin1]{inputenc} 
\usepackage{calc}         
\usepackage{graphicx}     
\usepackage{ifthen}       
\usepackage{pst-all}      
\usepackage{pst-poly}     
\usepackage{multido}      
\usepackage{pstricks-add} 
\usepackage{ulem} 

\newcommand{\ff}{\overrightarrow{\textit{ff}}}
\newcommand{\ffc}{\overleftarrow{\textit{ff}}}
\newcommand{\tto}{\overrightarrow{\textit{tt}}}
\newcommand{\ttc}{\overleftarrow{\textit{tt}}}
\newcommand{\tf}{\overrightarrow{\textit{tf}}}
\newcommand{\ft}{\overrightarrow{\textit{ft}}}
\newcommand{\ftc}{\overleftarrow{\textit{ft}}}
\newcommand{\tfc}{\overleftarrow{\textit{tf}}}

\newcommand{\FF}{\ {\textit{FF}}}
\newcommand{\TT}{\ {\textit{TT}}}
\newcommand{\TF}{\ {\textit{TF}}}
\newcommand{\FT}{\ {\textit{FT}}}

\newcommand{\FFc}{\overleftarrow{\textit{FF}}}
\newcommand{\TTc}{\overleftarrow{\textit{TT}}}
\newcommand{\TFc}{\overleftarrow{\textit{TF}}}
\newcommand{\FTc}{\overleftarrow{\textit{FT}}}

\newcommand{\FFo}{\overrightarrow{\textit{FF}}}
\newcommand{\TTo}{\overrightarrow{\textit{TT}}}
\newcommand{\TFo}{\overrightarrow{\textit{TF}}}
\newcommand{\FTo}{\overrightarrow{\textit{FT}}}

\newcommand{\f}{$\mathcal{F}(DA)\,$}

\newcommand{\shu}{\text{\rotatebox[origin=c]{-90}{$\exists$}}}

\begin{document}

\title{Deque languages,  automata and planar graphs}
\author{Stefano {Crespi Reghizzi} 
\and 
Pierluigi {San Pietro}
}

\institute{Dipartimento di Elettronica, Informazione e Bioingegneria (DEIB) and CNR - IEIIT\\
Politecnico di Milano, Piazza Leonardo da Vinci 32, 
Milano  I-20133\\
\email{stefano.crespireghizzi@polimi.it \quad pierluigi.sanpietro@polimi.it}}
\maketitle              

\normalem

\begin{abstract}
The memory of a \emph{deque} automaton  is more general than a queue or two stacks; to avoid overgeneralization, we consider quasi-real-time operation. Normal forms of such automata are given.   Deque languages form an AFL but not a full one. 
We define the  characteristic deque language, CDL, which combines  Dyck and AntiDyck (or FIFO) languages, and  homomorphically characterizes the deque languages. The notion of deque graph, from graph theory, well represents  deque computation by means of a planar hamiltonian graph on a cylinder, with edges visualizing producer-consumer relations for deque symbols. We give equivalent definitions of CDL by labelled deque graphs, by cancellation rules, and by means of shuffle and intersection of simpler languages. The labeled deque graph of a sentence generalizes traditional syntax trees. 
The  layout of deque computations on a cylinder is remindful of 3D models used in theoretical (bio)chemistry.
\end{abstract}

\pagenumbering{arabic}
\section{Introduction} 
This research pertains to the classical  investigations on  languages recognized by automata equipped with various types of auxiliary memory, such as pushdown stacks, queues, and combinations thereof.  Introduced by D. Knuth~\cite{DBLP:books/aw/Knuth68}, the \emph{deque} data-type is  common in computer science, where it is typically implemented by means of a bidirectional buffer.  
A deque memory  combines the standard operations of a queue and of two stacks but, unlike such simpler cases which impose serialization, it permits parallel execution or interleaving of some operations. 
Work on multihead/multi-tape Turing machines has also studied  simulation of deques, stacks and queues, e.g, in ~\cite{jacm/LeongSeiferas81}, where a deque is simulated in realtime by a  machine with four single head tapes; see also~\cite{DBLP:conf/cocoon/Petersen01} for deque simulation using stacks. On the other hand,  the deque has been rarely studied as a formal language model: the only   deque automaton model we know of is in~\cite{DBLP:journals/tcs/Ayers85}, but it is restricted and contains errors corrected in~\cite{DBLP:journals/tcs/Brandenburg87}.
\par
First, the present study of deque automata is an attempt to fill such gap and to establish formal relations especially with the family of queue automata~\cite{DBLP:journals/computing/Vollmar70,DBLP:journals/tcs/VauquelinF80,DBLP:journals/tcs/CherubiniCCM91}, which has recently attracted renewed interest (a survey is in~\cite{KutribMalcherW2018}).
A second reason to investigate deque languages is that they seem to fit,  better than traditional automata, with the linguistic models proposed by molecular biology and chemistry, to study spatial arrangements of macromolecule sequences. Such fitting is suggested by a similarity of representation, next outlined.  In their investigation of plane drawings of graphs on cylinder surfaces, Auer \emph{et al}.~\cite{DBLP:books/daglib/0035667,DBLP:conf/gd/AuerBBBG10} showed that ``a plane drawing is possible if, and only if, the graph is a \ldots deque  graph, i.e., the vertices of the graph can be processed according to a linear order and the edges correspond to items in the deque inserted and removed  at their end vertices''.
Similar  more complex embeddings of planar graphs on a cylinder are  considered in natural sciences, e.g., in~\cite{barthel2015toroidal,Searls2002} for chemistry and for RNA.
\par
Paper content and contributions. 
Sect.~\ref{sect:basicDef}  defines deque automata   (DA) consistently with existing definitions of pushdown and queue automata (PDA, QA). 
To prevent spontaneous moves to turn the deque without reading from input, thus simulating a Turing machine, we focus on  quasi-real-time (QRT) operations as in QA~\cite{DBLP:journals/tcs/CherubiniCCM91}. 
Our DA definition is rather robust, as the corresponding language family, denoted as \f, is unaffected by some typical variations: number of states, number of deque symbols processed per move, and restriction to real-time computations.
The family \f forms an abstract family of languages (AFL) but not a full one.
We define a convenient normal form 
using distinct tape alphabets for  operations on stacks and on queues and performing at most one deque operation per move.

To illustrate expressivity, we introduce a  family of languages featuring any number and ordering  of reversed and directed replications.
Sect.~\ref{sect:CharactDequeLang}   defines the {\em characteristic deque language} (CDL) which plays the role of the Dyck  and  the AntiDyck (a.k.a. FIFO) language, resp. for CF and queue languages and yields a Chomsky-Sch{\"u}tzenberger theorem for \f.
Sect.~\ref{sect:AuerGraphs}  exploits the   planar cylindrical graphs recently defined in~\cite{DBLP:conf/gd/AuerBBBG10} and develops them into a technique for analyzing and visualizing deque automata and languages. 
We prove that CDL computations on a DA are exactly represented by a \emph{labeled deque graph} (LDG). Then we characterize CDL by cancellation rules, unifying the classical rules for Dyck and AntiDyck languages. 
A closed formula can express a CDL as the shuffle of two CF languages intersected with a queue language;  its corollary for the whole \f family immediately follows. 
We end by showing how to extend the deque graph representing CDL words to all DA languages, thus endowing this family with a sort of structured syntax visualization. 
The conclusion  mentions directions for future research.

\section{Basic definitions and properties}\label{sect:basicDef}
A double ended  queue or \emph{deque} 
is an unbounded tape  containing a possibly empty string of symbols from an alphabet $\Gamma$. In a horizontal layout, the left  
end and the right end  of the tape are resp. called \emph{front} 
and \emph{tail},  both equipped with a reading and writing head.  The four deque \emph{operation types} are: write/read a symbol at  front, write/read a symbol at tail. Each reading operation cancels the symbol.
After each writing (resp. reading) operation at front,  the front head  moves left (resp. right) by one case; the tail head resp. moves right/left after writing/reading.
The deque is empty 
when the front and tail heads coincide.
\par
A \emph{deque automaton} $M$ is a nondeterministic  acceptor with one-way read-only input tape, finite-state control, and  deque memory tape, which initially is empty. 
$M$ is restricted to operate in QRT. 
\begin{definition}\label{Def:DA} A QRT \emph{deque automaton} ({\em DA})  is a 6-tuple
 $M=(\Sigma, \Gamma, Q, q_0, \delta, F)$ where $\Sigma$ is the terminal alphabet, $\Gamma$ the finite deque alphabet,  $q_0\in Q$ and $F \subseteq Q$ resp. the initial and final states; and  the transition relation $\delta$ is a finite subset: 
\begin{center}
$
\delta\, \subseteq\, \left( Q \times (\Sigma \cup \varepsilon)  \times \Gamma^*\times  \Gamma^* \times Q \times \Gamma^*\times  \Gamma^* \right) 
$
\end{center}
\end{definition}
For each 7-tuple, called a {\em transition}, $(q, a, \gamma_l, \gamma_r, q', \sigma,\rho)\in\delta$,  $q$ and $q'$ are resp. the present and next state, $\gamma_l, \sigma$ are the words resp. read and written on deque front, and  $\gamma_r, \rho$ are the words resp. read and written on deque tail. If $a=\varepsilon$ then the tuple is called an $\varepsilon$-transition.

A configuration is a 3-tuple  $(q, u, \gamma)\in Q\times \Sigma^*\times\Gamma^*$. 
A move is a binary relation $\to_M$ on configurations defined as follows.
We say that $(q, a u, \gamma_l\gamma\gamma_r)\to_M(q', u, \sigma\gamma\rho)$ if $ q,q' \in Q$, $a \in \Sigma\cup\varepsilon$, $u\in\Sigma^*$, 
$\sigma,\rho, \gamma,\gamma_l,\gamma_r\in\Gamma^*$ 
and the transition $(q,a,\gamma_l,\gamma_r,q',\sigma,\rho) \in \delta$. A move where $a=\varepsilon$ is called \emph{spontaneous}.

We assume that every DA $M$ operates in \emph{quasi-real-time} (QRT) with a delay $p\ge 1$. This means that every sequence of $p$ moves shifts the input tape of at least one position. 
We say that $M$ operates in \emph{real time} (RT) if $p=1$.

$M$ starts in the initial configuration $(q_0, w, \varepsilon)$ and accepts $w$ if there is a sequence of moves ending in a configuration $(q, \varepsilon, \varepsilon)$ with $q \in F$, 
i.e, $M$ scans from left to right the input word  starting with an empty tape and accepts upon reaching a final state provided  the tape is empty. 

\par
The language recognized by $M$, denoted by $L(M)$, is the set of words that $M$ accepts. The family of languages recognized by a DA is denoted  by \f. 
\par 
Clearly, restricted types of DA  correspond to QRT automata  having one queue, i.e., a FIFO tape, or two stacks, i.e., two LIFO tapes with a common bottom. 
We illustrate  Def.~\ref{Def:DA} with some basic languages.
\begin{example}\label{ex:dequelang}
Any CF language can be implemented on a DA using either the front or the tail head, 
since the QRT condition does not restrict the recognition capability  of nondeterministic PDA. 
We show in Fig.~\ref{fig:DequeAutomataEx1} (left) a DA using the front stack
 for the language of palindromes
$L_{pal}= \{u \,u^R \mid u \in \{a,b\}^+\}$. 
Then, since the front and tail stack operations are mutually independent, 
it is straightforward to define a DA accepting the intersection or the shuffle product ( $\shu \,$)  of two context-free languages.
\begin{figure}[h!]
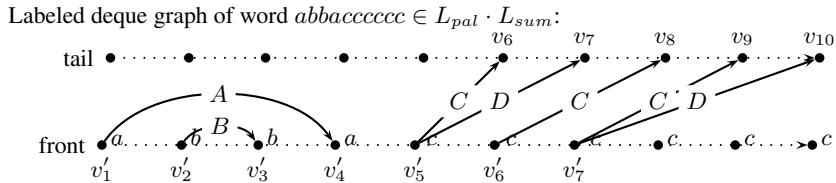

\begin{tabular}{c|c}
$
\begin{array}{l}
\text{DA for } L_{pal}= \{u \,u^R \mid u \in \{a,b\}^+\}:
\\
(q_0, a, \varepsilon, \varepsilon, q_0, A, \varepsilon), \,
(q_0, b, \varepsilon, \varepsilon, q_0, B, \varepsilon), \,
\\
(q_0, a, A, \varepsilon, q_1, \varepsilon, \varepsilon), \,
(q_0, b, B, \varepsilon, q_1, \varepsilon, \varepsilon), \,
\\
(q_1, a, A, \varepsilon, q_1, \varepsilon, \varepsilon), \,
(q_1, b, B, \varepsilon, q_1, \varepsilon, \varepsilon).
\\
\text{where $q_0$ initial  and  $q_1$  final.}
\end{array}
$
&
$
\begin{array}{l}\text{DA for } L_{sum}=\{c^\frac{n(n+1)}{2} \mid n \geq 1 \}:
\\
(p_0, c, \varepsilon, \varepsilon, p_1, D\,C, \varepsilon), 
(p_0, c, \varepsilon, \varepsilon, p_3, \varepsilon,  \varepsilon), 
\\
(p_1, c,\varepsilon,  C, p_2, C, \varepsilon),\,
(p_1, c, \varepsilon, C, p_3, \varepsilon, \varepsilon),
\\
(p_2, c, \varepsilon, D, p_1,  D\,C , \varepsilon),
(p_2, c, \varepsilon, C, p_2,  C , \varepsilon),
\\
(p_3, c, \varepsilon, C , p_3,\varepsilon,    \varepsilon)
(p_3, \varepsilon, \varepsilon, D , p_3,\varepsilon,    \varepsilon)

\\
\text{where $p_0$ initial and $p_3$ final.}
\end{array}
$
\end{tabular}
\\
\\
Labeled deque graph of word $abba cccccc \in  L_{pal}\cdot L_{sum}$:
\begin{center}
\scalebox{1.0}{%
\psset{arrows=->,labelsep=3pt,colsep=20pt,rowsep=22pt,nodealign=true}
\begin{psmatrix}
\dotnode{V1}&\dotnode{V2}&\dotnode{V3}&\dotnode{V4}&\dotnode{V5}&\dotnode{V6}\nput{90}{V6}{$v_6$}&\dotnode{V7}\nput{90}{V7}{$v_7$}
&\dotnode{V8}\nput{90}{V8}{$v_8$}&\dotnode{V9}\nput{90}{V9}{$v_9$}
&\dotnode{V10}\nput{90}{V10}{$v_{10}$}
\\
\dotnode{v'1}{$\,a$}\nput{-90}{v'1}{$v'_1$} &\dotnode{v'2}{$\,b$}\nput{-90}{v'2}{$v'_2$}
&\dotnode{v'3}{$\,b$}\nput{-90}{v'3}{$v'_3$}&\dotnode{v'4}{$\,a$}\nput{-90}{v'4}{$v'_4$}
&\dotnode{v'5}{$\;c$}\nput{-90}{v'5}{$v'_5$}&\dotnode{v'6}{$\;c$}\nput{-90}{v'6}{$v'_6$}
&\dotnode{v'7}{$\;\,c$}\nput{-90}{v'7}{$v'_7$}
&\dotnode{v'8}{$\,c$}&\dotnode{v'9}{$\,c$}&\dotnode{v'10}{$\,c$}
\nput[rot=0]{180}{V1}{tail\,}
\nput[rot=0]{180}{v'1}{front\,}

\ncarc[arcangle=55]{v'1}{v'4}\ncput*{$A$}
\ncarc[arcangle=55]{v'2}{v'3}\ncput*{$B$}
\ncline{v'5}{V6}\ncput*{$C$} 
\ncline{v'5}{V7} \ncput*{$D$}
\ncline{v'6}{V8}\ncput*{$C$} 
\ncline{v'7}{V9}\ncput*{$C$} 
\ncline{v'7}{V10}\ncput*{$D$} 
\ncline[linestyle=dotted]{V1}{V10} 
\ncline[linestyle=dotted]{v'1}{v'10} 
\end{psmatrix}
}
\end{center}
 \caption{Top: DAs of Ex.~\ref{ex:dequelang}. Bottom: labeled deque graph of a computation (explained in Sect.~\ref{ssect:LDGforDA}). }
 \label{fig:DequeAutomataEx1}
\end{figure}
Moreover, any language accepted by a QRT QA (model named $\text{NQA}_1$   in~\cite{DBLP:journals/tcs/CherubiniCCM91}) is obviously 
accepted by a DA that implements queue operations either in the direction front-to-tail ({\em ft}) or tail-to-front ({\em tf}). Incidentally, we observe that queue operations having opposite directions are mutually exclusive. 
The  nonsemilinear language  $L_{sum}=\{c^\frac{n(n+1)}{2} \mid n \geq 1 \}$ is recognized by the  DA in Fig.~\ref{fig:DequeAutomataEx1} (right)  implementing an \emph{ft} queue, similarly to the QAs in~\cite{DBLP:journals/tcs/CherubiniCCM91} (Stat. 3.9),  which define unary languages that are the solutions of a linear homogeneous difference equation. 

Then, the concatenation $L_{pal}\,L_{sum}$ is easily recognized by a DA that starts as the recognizer of $L_{pal}$ then, upon reading  $c$, switches to the  transition relation of $L_{sum}$. 
Notice that if the deque is not empty when scanning the first $c$, the recognizer of 
$L_{pal}\,L_{num}$  rejects the input.
Clearly,  
$L_{pal}\,L_{sum}$ 
cannot be recognized  by a nondeterministic QRT queue automaton~\cite{DBLP:journals/tcs/CherubiniCCM91} (which cannot define the language of palindromes).

At last,  a \emph{bordered} language~\cite{DBLP:journals/acta/KariK17} such as 
$\{u\, L_{pal}\, u \mid u \in \{a,b\}^+\}$ is easily recognized: the DA  non-deterministically stores the left border in the {\em tf} queue, then recognizes a word of $L_{pal}$ using  the front stack, then checks that the right border is identical to the stored border. Related  examples inspired  by 
the theory of DNA words~\cite{Searls2002} are possible.
\end{example}
From the previous examples and from the properties of QRT queue automata~\cite{DBLP:journals/tcs/CherubiniCCM91} we immediately have the following inclusions. 
\begin{theorem} Family \f  
strictly includes   the families of languages 
recognized by QRT queue automata and by QRT double pushdown automata.  
\end{theorem}
{\bf Remarks.}   QRT double pushdown automata can easily define   
the shuffle product  and the intersection of two CF languages.
The shuffle product  of two words contains all their interleavings, and, in the theory of parallel processes, represents 
 all possible total orderings (i.e., serial scheduling) of two independent sequences of events. In our setting, such independent processes are the PDAs operating on  front and on  tail. In fact, a deque, viewed as a memory device,  permits parallel execution of some operations, which are independent. This is in contrast with the  serial behavior of, say, a PDA.
\par
Several  variants of  Def.~\ref{Def:DA} are possible, differing with respect to   number of states, acceptance mode,  QRT constant value,  richness of operations per move, and in the structure of the tape alphabet. The following variants  have the same power. 
\begin{lemma}\label{lm:normalForms}
Let $M=(\Sigma, \Gamma, Q, q_0, \delta, F)$ be a DA  with delay $p\ge 1$ as in Def.~\ref{Def:DA}.
\begin{enumerate}
\item{\bf realtime.} There exists an  RT DA equivalent to $M$.
\item{\bf stateless.} There exists a DA with one state recognizing $L(M)\cup \varepsilon$.
\item{\bf simple.} There exists a so-called {\em simple} DA $M'$ equivalent to $M$,  
such that all its moves have the form 
$(q,a,\gamma_1,\gamma_2,q',\gamma_3,\gamma_4)$ (with $a \in \Sigma\cup \varepsilon)$,
where for $i = 1,2,3,4$: $\gamma_i \in \Gamma \cup \varepsilon$  and at most one $\gamma_i$ is not $\varepsilon$. 
\item{\bf partitioned.} There exists a  DA 
$M'$, 
equivalent to $M$ that uses four disjoint tape symbol alphabets, 
$\Gamma_{ff}$, $\Gamma_{tt}$, $\Gamma_{ft}$ and $\Gamma_{tf}$ 
resp. for operations on front stack (i.e., from front to front), tail stack (i.e., from tail to tail), front-to-tail queue, and tail-to-front queue,. i.e., 
for every move $(q,a,\gamma_1,\gamma_2,q',\gamma_3,\gamma_4)$ of $M'$, we have: 
\[\gamma_1\subseteq (\Gamma_{ff} \cup \Gamma_{tf})^*, \gamma_2\subseteq (\Gamma_{ft} \cup \Gamma_{tt})^*,  
\gamma_3\subseteq (\Gamma_{ff} \cup \Gamma_{ft})^*, \textit{ and } \gamma_4\subseteq (\Gamma_{tf} \cup \Gamma_{tt})^*. 
\]
\item $M$ may have $\varepsilon$ moves testing the deque for emptiness without increasing DA's expressive power.
\end{enumerate}
\end{lemma}
The  proofs are standard and are in the Appendix.
Space prevents detailed analysis of which  normal forms can be combined
together, and  discussion of other accepting modes, such as acceptance by final state only. 
In the following, we are mainly interested in deque automata that  combine the simple and the partitioned forms -- they are called 
\emph{SDA}. An example is in Def.~\ref{def:CharDequeLangAutomaton}.
\par
Since at most one symbol is  read or written in each move of an SDA, 
 a  shorthand  for specifying  moves (same in~\cite{DBLP:journals/tcs/VauquelinF80} for queue automata) is convenient.  
We define two copies $\overrightarrow{\Gamma}$, $\overleftarrow{\Gamma}$ of $\Gamma$ as follows: 
for every  symbol $A\in \Gamma$, a right (resp. left) pointing arrow $\overrightarrow{A}\in \overrightarrow{\Gamma}$, 
(resp. $\overleftarrow{A}\in\overleftarrow{\Gamma}$) means the symbol is written on the deque (resp. read from the deque);
the subset of $\Gamma= \Gamma_{ff}\cup \Gamma_{ft}\cup\Gamma_{tf}\cup\Gamma_{tt}$ uniquely identifies 
whether the operation  occurs at  front or at  tail. Therefore, the transition relation 
$\delta$ may be defined in short form as a subset of $Q \times \Sigma \times Q \times (\overrightarrow{\Gamma}\cup\overleftarrow{\Gamma})$.
\par
For instance, the move $(q,a,\varepsilon,\varepsilon,q',A_{ft},\varepsilon)$, which writes symbol  $A_{ft}$ on the front,
is now shortened to
 $(q, a, q', \overrightarrow{A_{ft}})$.
Similarly, 
 $(q, a, q', \overleftarrow{A_{tf}})$, which reads $A_{tf}$ from the front,  stands for
$(q, a, A_{tf},\varepsilon, q',\varepsilon,\varepsilon)$, and 
 $(q, a, q', \overrightarrow{A_{tf}})$  is $(q, a, A_{tf},\varepsilon, q',\varepsilon,\varepsilon)$, i.e., $A_{ft}$ is written to the front.
Then,  the transitions in short form of the DA for language 
$L_{pal}$ 
 in Fig.~\ref{fig:DequeAutomataEx1} are, without ambiguity:  
\\
$
(q_0, a, q_0, \overrightarrow{A}),\,
(q_0, b, q_0,\overrightarrow{B}),\,
(q_0, a, q_1,\overleftarrow{A}), \,
(q_0, b, q_1,\overleftarrow{B}),\,
(q_1, a, q_1,\overleftarrow{A}), \,
(q_1, b, q_1,\overleftarrow{B})
$.
\\
 Since the DA uses the front stack, the symbols  are in   $\Gamma_{ff}$ and should be renamed  $A\equiv A_{ff}$ and $B\equiv B_{ff}$.

\paragraph{Closure properties of \f.}
\begin{proposition}\label{prop-closureHomRevhom}
The family \f is an AFL, but it is not a full AFL. 
\end{proposition}
The proof is  standard and is in the Appendix. Thus, family \f enjoys a rich set of properties (e.g., closure under union, intersection with regular languages, 
nonerasing homomorphism, reverse homomorphism, Kleene's star, concatenation,  $\varepsilon$-free 
rational transduction and  inverse rational transduction), but it is not closed under erasing homomorphism.
Another useful property of any AFL 
is closure under $p$-limited erasing: 
\begin{definition}\label{def:locErasingHomom}\cite{HoUl79}
Given a language $L\subseteq\Sigma^*$, a homomorphism $h$ with the property that $h$ never maps more than $p$ consecutive symbols of any
sentence $x\in L$ to $\varepsilon$ is called a $p$\emph{-limited erasing}  on $L$.
A class of languages is said to be closed under $p$-limited
erasing if, for any language $L$ of the class and any $p$-limited
erasing $h$ on $L$, $h(L)$ is in the class.  
\end{definition}
When a homomorphism $h$ is a $p$-limited erasing on $L$, its effect is that, when applied to any word $w$ of  $L$, none of the factors of $w$ of length $p$ is entirely erased.

\subsection{Characteristic deque language and homomorphic characterization}\label{sect:CharactDequeLang}
This section introduces the language that is to DA as Dyck and AntiDyck languages resp. are to PDA and to QA.
The idea is that, as for simpler data structures, deque operations can be made into a terminal alphabet containing distinct copies of each operation.
Define for every $k\ge 1$, the finite alphabet $\Delta_k$ and some subsets thereof, as follows.
\begin{center}\renewcommand{\arraystretch}{1.5}
\begin{tabular}{l|lll|lll|lll|llll}
& \multicolumn{3}{c|}{$\overrightarrow{\Delta}_{ff,k}$} & \multicolumn{3}{c|}{$\overrightarrow{\Delta}_{ft,k}$} & \multicolumn{3}{c|}{$\overrightarrow{\Delta}_{tf,k}$}& \multicolumn{3}{c}{$\overrightarrow{\Delta}_{tt,k}$} &
\\\hline
$\overrightarrow{\Delta}_k $ & $\ff_1,$&  $\ldots,$ & $\ff_k$ & 
                  $\ft_1,$ &  $\ldots,$ & $\ft_k$ &
                  $\tf_1,$&  $\ldots,$ & $\tf_k$ & 
                  $\tto_1,$&  $\ldots,$ & $\tto_k$ &                
\\\hline\hline 
$\overleftarrow{\Delta}_k $ & $\ffc_1,$&  $\ldots,$ & $\ffc_k$ &
                  $\ftc_1,$ &  $\ldots,$ & $\ftc_k$ &
                  $\tfc_1,$&  $\ldots,$ & $\tfc_k$ & 
                  $\ttc_1,$&  $\ldots,$ & $\ttc_k$ & 
\\\hline
& \multicolumn{3}{c|}{$\overleftarrow{\Delta}_{ff,k}$} & \multicolumn{3}{c|}{$\overleftarrow{\Delta}_{ft,k}$} & \multicolumn{3}{c|}{$\overleftarrow{\Delta}_{tf,k}$}& \multicolumn{3}{c}{$\overleftarrow{\Delta}_{tt,k}$}&                                            
\end{tabular}
\end{center}
The alphabet is $\Delta_k = \{\ff_1,\ldots,\ttc_k \}$. Thus, the set of ``open brackets" in $\Delta_k$ is denoted by $\overrightarrow{\Delta_k}$, the set of closing brackets of the form $\ffc_j$ as $\overleftarrow{\Delta}_{ff,k}$, etc.

\par
The  natural definition of  CDL is by means of a deterministic RT DA with just one state, in analogy with the PDA recognizing the Dyck language.
\begin{definition}\label{def:CharDequeLangAutomaton}
For each $k\geq 1$, the \emph{characteristic deque language} (CDL), denoted by $DQ_k \subset \Delta_k^*$, is the language accepted by the  SDA
\[
A_{CDL_k}=\left(\{q_0\}, \Delta_k, \Gamma,\delta, q_0, \{q_0\}\right)
\] 
where $\Gamma =\bigcup_{1 \leq j \leq k} (\{\FF_j,  \TF_j, 
\TT_j,  \FT_j\})$ and 
the transition relation is defined as follows (the only state of $A_{CDL_k}$ is omitted): 
\\\centerline{ $
\delta =\bigcup_{1 \leq j \leq k} \left\{ 
\begin{array}{l}
(\ff_j,\, \FFo_j), (\ft_j,\FTo_j), (\tf_j,\, \TFo_j),(\tto_j,\, \TTo_j),  
\\
(\ffc_j,\, \FFc_j), (\ftc_j,\FTc_j), (\tfc_j,\, \TFc_j),(\ttc_j,\, \TTc_j)  
\end{array}
\right\}.
$}
\end{definition}
Notice that this machine is deterministic, real time and   ``stateless''. 
To illustrate, word 
$
\tto_1  \ff_1 \tto_2 \ffc_1 \ft_1 \ttc_2 \ft_2 \ttc_1 \ftc1 \ftc_2 \in DQ_2
$ 
is accepted with the computation:
\\
$
\begin{aligned}
\varepsilon &\stackrel {\tto_1}\Longrightarrow  \TT_1 
\stackrel {\ff_1} \Longrightarrow  \FF_1 \TT_1   
\stackrel {\tto_2} \Longrightarrow  \FF_1 \TT_1 \TT_2   
\stackrel {\ffc_1 }\Longrightarrow  \TT_1 \TT_2 
\stackrel {\ft_1 }\Longrightarrow \TT_1 \TT_2 \FT_1 \\
&\stackrel {\ttc_1 }\Longrightarrow  \TT_2 \FT_1 
\stackrel {\ft_2 }\Longrightarrow  \TT_2 \FT_1 \FT_2
\stackrel {\ttc_2}\Longrightarrow  \FT_1 \FT_2 
\stackrel {\ftc_1 }\Longrightarrow   \FT_2
\stackrel {\ftc_2 }\Longrightarrow \varepsilon 
\end{aligned}
$
\\
In words, the sentences of  CDL are the  sequences  that obey the natural schedule of deque operations.  Other characterizations of CDL in terms of graphs, cancellation rules and shuffles
will be given in Sect.~\ref{sect:AuerGraphs}.

Using CDL we characterize deque languages $\grave{a} \emph{ la}$   Chomsky-Sch{\"u}tzenberger.

\begin{theorem}\label{theor:homomCharacteriz}
A language $L \subseteq \Sigma^* $ is accepted by a  deque automaton if, and only if, 
there exist $k>0,\,p>0$, a finite  alphabet $\Theta$, a homomorphism $g: \Theta \to \Delta_k^* $,
a regular language $R$ on  $\Theta$, and a $p$-limited erasing  $h: \Theta \to \Sigma\cup \varepsilon$ on $R$ such that
$
L = h\left(g^{-1}(DQ_k) \cap R\right).
$
\end{theorem}
\proof
By the  closure properties of an AFL it is obvious that if $L =h\left(g^{-1}(DQ_k) \cap R\right)$ for some $h,g,R,k$ verifying the statement of the theorem, then $L$ can be 
recognized by a DA.
Let now $L$ be recognized by a DA $M= \left(Q, \Sigma,  \Gamma, \delta, q_0, F  \right)$, which we assume to be in simple partitioned normal form.
Therefore, $\delta\subseteq Q \times (\Sigma\cup \varepsilon)\times Q \times (\overrightarrow{\Gamma}\cup\overleftarrow{\Gamma})$, where
$\overrightarrow{\Gamma}$ and  $\overleftarrow{\Gamma}$ are two disjoint copies
of the tape alphabet $\Gamma$.
\\
Since $\Gamma$ is partitioned in $\Gamma_{ft}, \,\Gamma_{tt}$, etc., the alphabet $\overrightarrow{\Gamma}\cup\overleftarrow{\Gamma}$
can be considered as a 
characteristic alphabet $\Delta_k$, with $k=|\Gamma|$. 
Define the finite alphabet $\Theta=\delta$, i.e., $\Theta$ is the set of all quadruples in $\delta$.
Let $g:\Theta\to\Delta_k$ be the homomorphism defined by $g(\langle q,a,q',A\rangle)= A$, 
for every $q,q'\in Q, a\in\Sigma\cup \varepsilon$, $A \in \overrightarrow{\Gamma}\cup\overleftarrow{\Gamma}$. 
Let $R$ be the local language in $\Theta^*$ defined by the pairs of consecutive transitions of $M$ 
(e.g.,  $\langle q,a,q',\overleftarrow{A}\rangle\langle q',b,q'',\overleftarrow{B}\rangle$, etc.)
\\
Let  $h$ be
the projection of $\Theta$ on 
the 2nd component, i.e., $h(\langle q,a,q',A\rangle)=a$. 
It is obvious that $h\left(g^{-1}(DQ_k) \cap R\right)$ is $L$.
Homomorphism $h$ is $p$-erasing for $R$ and thus for the language $g^{-1}(DQ_k) \cap R$, where $p$ is the QRT constant of $M$, since it 
returns $\varepsilon$ only in correspondence to $\varepsilon$-transitions of $M$. 
\qed
By a standard procedure (for QA in~\cite{DBLP:journals/tcs/VauquelinF80}), in the statement of Theor.~\ref{theor:homomCharacteriz} 
it is possible to assume $k=2$. In fact, let $\rho: \Delta_k \to \Delta_2^+$ be the homomorphism defined by $\rho(a_j) =
a_1 a_2^j $, for every $a_1, a_2, a_j \in \overrightarrow{\Delta_k}$, and $\rho(a_j) =
a_2^j\, a_1$, for every $a_1, a_2, a_j \in \overleftarrow{\Delta_k}$. Then $\rho^{-1}(DQ_2) = DQ_k$. \qed

\subsection{Example of expressiveness  of deque languages}\label{sect:expessivenessAndRepresentation}
Deque automata have a  noteworthy capability to define languages that replicate,  any number of times,  a factor or its reversal.
We can introduce a schema for specifying replications, by means of a regular language $\Pi \subseteq \{D, R \}^*$ 
where $D$ and $R$ resp.  stand for ``direct'' and ``reverse''. 
Intuitively, a word such as $DDRD$ specifies that a given word $u\in \Sigma^+$ is followed by  4 replicas:  $uu u^R u$. 
We can define languages parameterized by a replication schema, e.g., $
L_{(D^* R DR)} = \left\{u\, u^*\, u^R\, u\, u^R \mid u \in \Sigma^+ \right\}
$.
A   family $\{L_{(\Pi)}\}$, of languages parameterized by a replication schema $\Pi$, called  \emph{regular replica family}, is defined as follows, using
the  following family of homomorphisms,  for every $u \in \Sigma^+$:   $\rho_u: \{D,R\}\to \Sigma^+$ is defined as $\rho_u(D)=u, \rho_u(R)=u^R$. Then the language is
\\ 
\centerline{$ L_{(\Pi)} = \left\{ w \in \Sigma^+ \mid \exists  u \in \Sigma^+, \pi \in \Pi : w = \rho_u(\pi)\right\}$.}
\begin{proposition}\label{prop-replicafamily}
For each regular replication schema $\Pi \subseteq \{D, R \}^*$, the language $L_{(\Pi)}$ is in \f. 
\end{proposition}
\proof (the complete proof is in the Appendix) Let $L = \{ w \in (\Sigma\cup\{D,R\})^+ \mid w = u(Du\cup Ru^R)^+, u \in \Sigma^+\}$.
We claim that $L$ is  in \f, hence also $L_{(\Pi)}$ is in \f:  let $\Pi'$ be the regular language 
obtained by the shuffle of $\Pi$ with  $\Sigma^+$; by closure under intersection
with regular languages also $L\cap \Pi'$ is in \f; by applying an obvious 1-limited erasing we obtain $L_{(\Pi)}$.
In the initial state $q_0$, given a word $uD\dots$ or $uR\dots$, $M$ stores factor $u$ in the deque at the tail's end, so that when reading 
from the front  the deque content is $u$, when reading from the tail it is $u^R$.
Upon reading $D$ or $R$, $M$ changes its state to the state $q_F$ or, resp., to the state $q_T$ and writes a new symbol $Z$ to the tail or, resp., to the front; the content of the deque 
is $uZ$ or, resp., $Zu$. 
In state $q_F$, $M$ compares the current input symbol $a$ with the symbol at the front and writes $a$ to the tail. 
When $D$ or $R$ are scanned and $Z$ is read from the front, $M$ changes its state to $q_F$ (if $D$) or $q_T$  (if $R$) 
and writes $Z$ to the tail or resp. to the front; the deque content is still $uZ$ or resp. $Zu$. 
The behavior in state $q_T$ is symmetrical: $M$ compares the input with $u^R$ read from the tail and rewrites it to the front, so that at the end the tape content is still $u$.
 $M$ guesses when the current factor is the last one and  then ceases to store the replica on the tape. \qed
 
\section{Characterization of deque sentences  by planar graphs}\label{sect:AuerGraphs} 
\begin{figure}[h!]
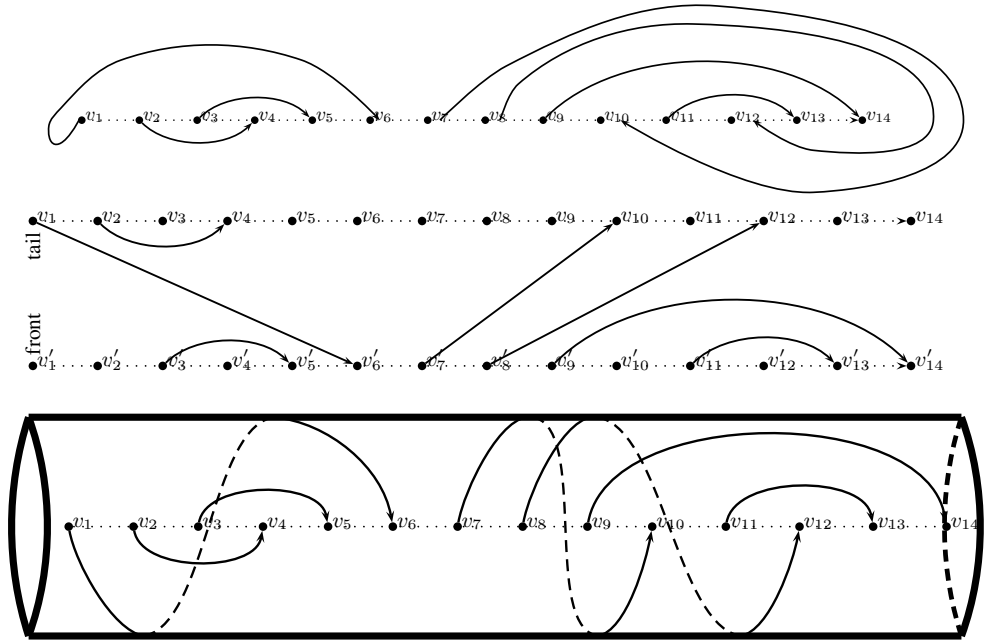

\vspace{0.8cm}
\begin{center}
 \scalebox{0.80}{%
\psset{arrows=->,labelsep=3pt,colsep=15pt,rowsep=80pt,nodealign=true}
\begin{psmatrix}
\dotnode{V1}{$v_1$}&\dotnode{V2}{$v_2$}&\dotnode{V3}{$v_3$}&\dotnode{V4}{$v_4$}&\dotnode{V5}{$v_5$}&\dotnode{V6}{$v_6$}&\dotnode{V7}{$v_7$}
&\dotnode{V8}{$v_8$}&\dotnode{V9}{$v_9$}&\dotnode{V10}{$v_{10}$}&\dotnode{V11}{$v_{11}$}
&\dotnode{V12}{$v_{12}$}&\dotnode{V13}{$v_{13}$}&\dotnode{V14}{$v_{14}$}
\\
\pscurve[showpoints=false](-0.2,3.2)(-0.6,2.8)(-0.7,3.1) (-0.5,3.4) (0.5,4.2) (3.5,4.2) (4.8,3.2)

\ncarc[arcangle=-45]{V2}{V4}

\ncarc[arcangle=45]{V3}{V5}
\pscurve[showpoints=false](5.8,3.2)(6.9,4.1)(10.4,5.1)(14.5,3.2) (12.0,2.0)(8.8,3.2)
\pscurve[showpoints=false](6.8,3.2)(7.1,3.8)(10.0,4.8)(14.0,3.2)(12.0,2.7)(11.0,3.2)
\ncarc[arcangle=45]{V9}{V14}
\ncarc[arcangle=45]{V11}{V13}
\ncline[linestyle=dotted]{V1}{V14} 
\end{psmatrix}
}
\end{center}
\begin{center}
\vspace{-2cm}
 \scalebox{0.90}{%
\psset{arrows=->,labelsep=3pt,colsep=15pt,rowsep=50pt,nodealign=true}
\begin{psmatrix}
\dotnode{V1}{$v_1$}&\dotnode{V2}{$v_2$}&\dotnode{V3}{$v_3$}&\dotnode{V4}{$v_4$}&\dotnode{V5}{$v_5$}&\dotnode{V6}{$v_6$}&\dotnode{V7}{$v_7$}
&\dotnode{V8}{$v_8$}&\dotnode{V9}{$v_9$}&\dotnode{V10}{$v_{10}$}&\dotnode{V11}{$v_{11}$}
&\dotnode{V12}{$v_{12}$}&\dotnode{V13}{$v_{13}$}&\dotnode{V14}{$v_{14}$}
\\
\dotnode{v'1}{$v'_1$}&\dotnode{v'2}{$v'_2$}&\dotnode{v'3}{$v'_3$}&\dotnode{v'4}{$v'_4$}&\dotnode{v'5}{$v'_5$}&\dotnode{v'6}{$v'_6$}&\dotnode{v'7}{$v'_7$}
&\dotnode{v'8}{$v'_8$}&\dotnode{v'9}{$v'_9$}&\dotnode{v'10}{$v'_{10}$}&\dotnode{v'11}{$v'_{11}$}
&\dotnode{v'12}{$v'_{12}$}&\dotnode{v'13}{$v'_{13}$}&\dotnode{v'14}{$v'_{14}$}

\nput[rot=90]{-90}{V1}{tail}
\nput[rot=90]{90}{v'1}{front}
\ncline{V1}{v'6}
\ncarc[arcangle=-45]{V2}{V4}

\ncarc[arcangle=45]{v'3}{v'5}
\ncline{v'7}{V10} 
\ncline{v'8}{V12} 
\ncarc[arcangle=45]{v'9}{v'14}
\ncarc[arcangle=45]{v'11}{v'13}
\ncline[linestyle=dotted]{V1}{V14} 
\ncline[linestyle=dotted]{v'1}{v'14} 
\end{psmatrix}
}
\\
\vspace{0.5cm}
\scalebox{0.90}{%
\psset{arrows=->,labelsep=3pt,colsep=15pt,rowsep=35pt,nodealign=true}
\begin{psmatrix}
\pnode{A1} & & & & \pnode{A2}& & & & \pnode{A3}& \pnode{A4}& & & & &\pnode{A15}
\\
&\dotnode{V1}{$v_1$}&\dotnode{V2}{$v_2$}&\dotnode{V3}{$v_3$}&\dotnode{V4}{$v_4$}&\dotnode{V5}{$v_5$}&\dotnode{V6}{$v_6$}&\dotnode{V7}{$v_7$}
&\dotnode{V8}{$v_8$}&\dotnode{V9}{$v_9$}&\dotnode{V10}{$v_{10}$}&\dotnode{V11}{$v_{11}$}
&\dotnode{V12}{$v_{12}$}&\dotnode{V13}{$v_{13}$}&\dotnode{V14}{$v_{14}$}
\\
\pnode{B1}& &\pnode{B2}&  & & & & & &\pnode{B3} & & \pnode{B4}& & &\pnode{B15}

\ncline[arrows=-, linewidth=3pt]{A1}{A15}
\ncline[arrows=-, linewidth=3pt]{B1}{B15}
\ncarc[arcangle=20, arrows=-, linewidth=3pt]{A1}{B1}
\ncarc[arcangle=-18, arrows=-, linewidth=3pt]{A1}{B1}
\ncarc[arcangle=20, arrows=-, linewidth=3pt]{A15}{B15}
\ncarc[arcangle=-18, arrows=-, linestyle=dashed,linewidth=2pt]{A15}{B15}
\ncline[arrows=-, linestyle=dotted]{V1}{V14}

\psset{arrows=->,linewidth=1pt}
\nccurve[arrows=-,angleA=-80,angleB=170,ncurv=0.5]{V1}{B2}

\nccurve[arrows=-,linewidth=1pt,linestyle=dashed,angleA=10,angleB=-170,ncurv=0.5]{B2}{A2}

\nccurve[angleA=-5,angleB=100,ncurv=0.7]{A2}{V6}

\ncarc[arcangle=-80]{V2}{V4}

\ncarc[arcangle=80]{V3}{V5}

\nccurve[arrows=-,angleA=80,angleB=170,ncurv=0.5]{V7}{A3}

\nccurve[arrows=-,linewidth=1pt,linestyle=dashed,angleA=-10,angleB=170,ncurv=0.5]{A3}{B3}

\nccurve[angleA=10,angleB=-100,ncurv=0.5]{B3}{V10}

\nccurve[arrows=-,angleA=80,angleB=170,ncurv=0.5]{V8}{A4}

\nccurve[arrows=-,linewidth=1pt,linestyle=dashed,angleA=-10,angleB=170,ncurv=0.5]{A4}{B4}

\nccurve[angleA=10,angleB=-100,ncurv=0.5]{B4}{V12}

\ncarc[arcangle=80]{V9}{V14}

\ncarc[arcangle=80]{V11}{V13}
\end{psmatrix}
}
\end{center}
\caption{ Top: planar graph with hamiltonian path dotted. Middle: the same graph represented as deque graph  with duplicated vertices. Bottom: 3D linear cylindric drawing of the same graph. The middle graph is  achieved by cutting along the hamiltonian path and unrolling the cylinder surface, and  duplicating vertices.}
\label{fig:cylindric2plane}
\end{figure}

 An insightful analysis of deque operation sequences has been recently obtained within research on graph drawing by Auer et al. \cite{DBLP:conf/gd/AuerBBBG10,DBLP:books/daglib/0035667}.
Given a planar graph $(V,E)$  having exactly one hamiltonian path, imagine to  draw the vertices $V$ so that the hamiltonian path edges, called $E_H$,  lay on a straight line, called a  \emph{linear layout}, which thus visualizes a total order on $V$. The other edges $E-E_H$ can be drawn without crossing.
The direction of  edge $(v_i, v_j)$, $i< j$,  is $v_i\to v_j$. Such a drawing is called a \emph{deque graph} because each non-hamiltonian edge  represents two operations that resp. insert and remove an item from the deque. Extending previous work on the linear layout of  simpler data structures (one/two pushdown stacks,  queue), Auer et al. \cite{DBLP:conf/gd/AuerBBBG10,DBLP:books/daglib/0035667} have precisely characterized the deque graphs.
We informally  introduce  the  concepts  from the cited studies that are relevant for studying deque automata and languages. 
\\\indent
A  deque graph  represents a sequence of deque operations: writing occurs at the origin $v_i$ of a (non-hamiltonian) edge $v_i\to v_j$ and reading at the destination  $v_j$. Since the  hamiltonian path   totally orders vertices, if we label each vertex with a terminal character, we  obtain a word. 
Fig. \ref{fig:cylindric2plane}, derived from \cite{DBLP:conf/gd/AuerBBBG10}, shows  (top)  a planar deque graph, (bottom) its 3D representation on a torus or cylinder, and (middle) its plane layout obtained by cutting the torus surface along the hamiltonian path and duplicating vertices.   
Conceptually, the primary representation in \cite{DBLP:conf/gd/AuerBBBG10,DBLP:books/daglib/0035667} is the graph  drawn on torus, called a \emph{linear cylindric drawing} (LCD). The main result is that an LCD represents a valid sequence  of deque operations,  and is called a \emph{deque graph}, if, and only if, it is a planar graph with one hamiltonian path. 
For practicality  the plane layout (Fig.~\ref{fig:cylindric2plane}, middle) is preferred, but an LCD 
would be more appropriate as a model of 3D-sequences, such as  (bio)chemical strings of molecules.
To transform an LCD into a  deque graph, the hamiltonian path is duplicated and each vertex $v_i$ is represented by a pair of points $v_i, v'_i$, resp. placed on the upper and lower path, which resp. represent the deque tail and front sides.    
We  now introduce a labeling of the vertices, to associate a deque graph with a word.
\begin{definition} Let $\Sigma$ be a terminal alphabet and let $G=(V,E)$ 
be  a deque graph. 
A \emph{labeled deque graph} (LDG)  is defined by the pair $(G, \lambda)$ where $\lambda$ is the \emph{labeling function} $\lambda: V \to \Sigma$, which assigns a terminal character to each vertex.
The \emph{word defined} by an LDG $(G, \lambda)$, denoted by $W(G, \lambda)$,  is $\lambda(v_1)\lambda(v_2) \ldots \lambda(v_n)$, where $v_1 v_2 \dots v_n$ is  the hamiltonian path.
Let  $\mathcal{G}=\{\ldots,(G_m, \lambda) , \ldots\}$ be a family of LDGs. The \emph{language defined by the graph family} is $W(\mathcal{G}) =\bigcup_{(G_m, \lambda_m)\in \mathcal{G}} W(G_m, \lambda_m)$.
\end{definition}

\paragraph{Definition of CDL by means of graphs.}   
Using LDGs, we study the sequences of deque operations which may occur in a CDL,  i.e., the sentences of language $DQ_k$ of Def.~\ref{def:CharDequeLangAutomaton}.
We only consider deque graphs such that exactly one deque operation occurs at each vertex, i.e., one non-hamiltonian edge impinges on it. 
Auer's model  allows any number of non-hamiltonian edges per vertex, but here
we want to ensure that the label of each vertex carries enough information about the type of the incoming/outgoing edge.  
\\
There are four edge types
depending on where their source and destination vertices lie on the unrolled cylinder: 
front-to-front ({\em ff}),  tail-to-tail ({\em tt}), front-to-tail ({\em ft}) and tail-to-front ({\em tf}). The type of an edge $e$ is denoted by $\tau(e)$.  Next, we label the vertices with "`brackets"' that carry the information about the type of edge they belong to. 

\begin{definition}\label{def-LDG}
Let $G=(V,E)$ be a deque graph. For each $k\geq 1$,we denote by  $\mathcal{G}_k$ the family of all $(G,\lambda)$, where 
 the \emph{characteristic  labeling function} $\lambda: V \to \Delta_k$ is defined as follows.
 For each edge  $e=(v_i \to v_j)\in E$, $\exists 1 \leq h \leq k$ and
\\
$
\begin{array}{l||l}
\text{if $\tau(e)=\textit{ff}$ then } \lambda(v_i)= \ff_h;\, \lambda(v_j)= \ffc_h
&
\text{if $\tau(e)=\textit{tt}$  then } \lambda(v_i)= \tto_h;\, \lambda(v_j)= \ttc_h
\\
\text{if $\tau(e)=\textit{ft}$  then } \lambda(v_i)= \ft_h;\, \lambda(v_j)= \ftc_h
&
\text{if $\tau(e)=\textit{ft}$ then } \lambda(v_i)= \tf_h;\, \lambda(v_j)= \tfc_h.
\end{array}
$
\end{definition}

\begin{example}\label{ex:DQ2graph}
Fig. \ref{fig:labeledDQ2graph}  show an LDG based on the characteristic labeling over $\Delta_2$ of the deque graph in Fig. \ref{fig:cylindric2plane} (middle).
\begin{figure}[h!]
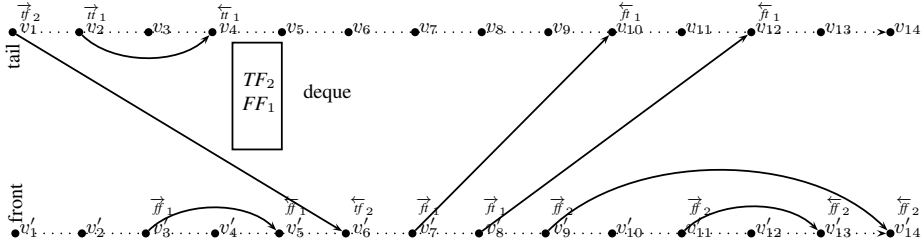

\begin{center}
\scalebox{0.85}{%
\psset{arrows=->,labelsep=3pt,colsep=15pt,rowsep=70pt,nodealign=true}
\begin{psmatrix}
\dotnode{V1}{$\stackrel{\tf_2}{v_1}$}&\dotnode{V2}{$\stackrel{\tto_1}{v_2}$}&\dotnode{V3}{$v_3$}&\dotnode{V4}{$\stackrel{\ttc_1}{v_4}$}&\dotnode{V5}{$v_5$}&\dotnode{V6}{$v_6$}&\dotnode{V7}{$v_7$}
&\dotnode{V8}{$v_8$}&\dotnode{V9}{$v_9$}
&\dotnode{V10}{$\stackrel{\ftc_1}{v_{10}}$}&\dotnode{V11}{$v_{11}$}
&\dotnode{V12}{$\stackrel{\ftc_1}{v_{12}}$}&\dotnode{V13}{$v_{13}$}&\dotnode{V14}{$v_{14}$}
\\
\dotnode{v'1}{$v'_1$}&\dotnode{v'2}{$v'_2$}
&\dotnode{v'3}{$\stackrel{\ff_1}{v'_3}$}&\dotnode{v'4}{$v'_4$}
&\dotnode{v'5}{$\stackrel{\ffc_1}{v'_5}$}&\dotnode{v'6}{$\stackrel{\tfc_2}{v'_6}$}
&\dotnode{v'7}{$\stackrel{\ft_1}{v'_7}$}
&\dotnode{v'8}{$\stackrel{\ft_1}{v'_8}$}&\dotnode{v'9}{$\stackrel{\ff_2}{v'_9}$}&\dotnode{v'10}{$v'_{10}$}&\dotnode{v'11}{$\stackrel{\ff_2}{v'_{11}}$}
&\dotnode{v'12}{$v'_{12}$}&\dotnode{v'13}{$\stackrel{\ffc_2}{v'_{13}}$}
&\dotnode{v'14}{$\stackrel{\ffc_2}{v'_{14}}$}
\psframe(-10.8,1.3)(-10,3)\rput(-10.4,2.2){$\begin{array}{c}\TF_2 \\ \FF_1\end{array}$}
\rput(-9.3,2.2){deque}
\nput[rot=90]{-90}{V1}{tail}
\nput[rot=90]{90}{v'1}{front}
\ncline{V1}{v'6}
\ncarc[arcangle=-45]{V2}{V4}

\ncarc[arcangle=45]{v'3}{v'5}
\ncline{v'7}{V10} 
\ncline{v'8}{V12} 
\ncarc[arcangle=45]{v'9}{v'14}
\ncarc[arcangle=45]{v'11}{v'13}
\ncline[linestyle=dotted]{V1}{V14} 
\ncline[linestyle=dotted]{v'1}{v'14} 
\end{psmatrix}
}
\end{center}
\caption{A characteristic labeling over $\Delta_2$ of the deque graph in  Fig.~\ref{fig:cylindric2plane} (middle).  
The deque content after  four steps is shown.}
\label{fig:labeledDQ2graph}
\end{figure}
The word defined by the graph is:\\
$ \tf_2  \tto_1 \ff_1   \ttc_1  \ffc_1 \tfc_2 \ft_1 \ft_1 \ff_2 \ftc_1\ff_1 \ftc_1 \ffc_1\ffc_2
$. 
The deque content after reading the prefix  $\tf_2 \tto_1 \ff_1 \ttc_1 $ is $\TF_2\FF_1$, which in Fig.~\ref{fig:labeledDQ2graph} is visible as the vertical cut of edges $(v_1, v'_6)$ and $(v'_3, v'_5)$. 
\end{example}
We show that each CDL is the language defined by a family of characteristic LDGs.

\begin{theorem}\label{th:DQeqLDG}
The following identity holds:    $W(\mathcal{G}_k) = DQ_k$.
\end{theorem}
The proof,  in the Appendix, 
first makes precise the definition of deque machines of~\cite{DBLP:books/daglib/0035667}, that we call \emph{Deque Graph Machines} (DGMs), and we enrich  with a finite alphabet to label vertices, not present in ~\cite{DBLP:books/daglib/0035667}, 
so that it can be seen as accepting words, in particular 
over the characteristic alphabet. Such machine is called a $\Delta_k$-DGM. 
When storing an edge $e=\langle \alpha, \beta\rangle$ of an LDG $(G,\lambda)$, with $G =(V,E)$, also the 
label of the  destination vertex $\beta$ must be stored; the label is denoted, with an abuse of notation,  
by $\lambda(e)$ (with $\lambda: E \to \overleftarrow{\Delta_k}$). 
\par
The result is proved by showing that, when reading the same prefix $y$ of a  word $yx$ in $DQ_k$,  the configurations $\langle x,q_0,u\rangle$ of $A_{CDL_k}$ and 
$\langle x,q_0,v\rangle$ of $\Delta_k$-DGM  are such that 
$\Lambda(v)=u$, where $\Lambda: E \to  \Gamma$ 
is the homomorphism defined as the ``upper case'' version of $\lambda$, i.e., 
if $\lambda(e) =\ffc_j$ then $\Lambda(e) = \FF_j$, etc.

\paragraph{Definition of characteristic deque languages via cancellation rules.}\label{sect:cancellationRules}
The well known definition of Dyck's languages as the equivalence class of the words obtained 
by applying the cancellation rule $( \, ) \to \varepsilon$ is now  combined with the similar definition of the AntiDyck languages \cite{DBLP:journals/tcs/VauquelinF80}  into a set of cancellation rules defining  $DQ_k$ languages. 

\begin{theorem}\label{theorCancellationRules}
A word over alphabet $\Delta_k$, $k\geq 1$, is in language $DQ_k$ if, and only if, 
the  application  zero or more times (in any order) of the following \emph{cancellation rules} (CR)  
reduces the word to empty.\emph{ For all} $v \in {\Delta_k}^*\,:$ 
\\
$\begin{array}{ll}
1. &\forall u \in (\Delta_k - \overleftarrow{\Delta}_{ff,k})^*, 
\forall x \in (\overleftarrow{\Delta}_{ft,k} \cup \overrightarrow{\Delta}_{tf,k} \cup \Delta_{tt,k})^* :
 u \,\ff_i \, x\, \ffc_i \, v \to u xv
\\
2. & \forall u \in (\Delta_k - \overleftarrow{\Delta}_{tt,k})^*, 
\forall x \in (\Delta_{ff,k}\cup \overrightarrow{\Delta}_{ft,k} \cup \overleftarrow{\Delta}_{tf,k} )^* :
 u \, \tto_i \, x\, \ttc_i\, v \to uxv
\\
3. &  \forall x \in (\overrightarrow{\Delta}_{ft,k})^* : 
\ft_i\,x \,\ftc_i \, v\to xv
\\
4. & \forall x \in (\overrightarrow{\Delta}_{tf,k})^* : 
\tf_i\,x \,\tfc_i\, v \to xv\,.
\end{array}
$
\end{theorem}
The proof, in the Appendix, shows that, for any word accepted by $A_{CDL_k}$,   each cancellation step preserves correctness, and conversely.

\paragraph{Composition of Dyck and AntiDyck.}
CRs can be applied in the following order, repeatedly scanning the input word  from left to right: 
CR 1 and 2 delete all  Dyck  symbols; then, CR 3 and 4 all AntiDyck
symbols. CR 1, 2 could be reformulated so that an application of either 
rule is not necessarily leftmost. 
Such remarks suggest to express  CDL as a suitable combination of Dyck and AntiDyck languages.
Define two CF languages and a \f language, using the Dyck languages $Dyck_{ff},\,Dyck_{tt}$, resp. over 
$\Delta_{ff,k}$ and  $\Delta_{tt,k}$, and the AntiDyck ($Adyck$) languages over $\Delta_{ft,k}$ and  $\Delta_{tf,k}$:
\begin{center} 
$
\begin{array} {l|l}
H_1= \left(Dyck_{ff}\cdot (\overrightarrow{\Delta}_{ft,k}  \cup \overleftarrow{\Delta}_{tf,k})^* \right)^* Dyck_{ff} 
&
H_3= \phi^{-1}\left((\textit{Adyck}_{ft} \cup \textit{Adyck}_{tf})^*   \right)
\\
H_2= \left(Dyck_{tt}\cdot (\overleftarrow{\Delta}_{ft,k}\cup \overrightarrow{\Delta}_{tf,k})^* \right)^*Dyck_{tt}
&
\text{with} \left\{
   \begin{array} {l}
   \phi(a)= \varepsilon,\, a \in \Delta_{ff,k} \cup \Delta_{tt,k}
   \\
   \phi(a)= a,\,   a \in \Delta_{ft,k} \cup \Delta_{tf,k}
   \end{array}
   \right.
\end{array}
$
\end{center}

\begin{theorem}\label{theor:formulaCDL}
For each CDL the identity holds: 
$
 DQ_k = (H_1\,   \shu \,  H_2 )  \cap H_3
$. 
\end{theorem}
\proof 
Let $H_4=(H_1\,   \shu\,  H_2 )  \cap H_3$.
Claim $H_4\subseteq DQ_k$ : we show that if $ w' \in H_4$, $w'$ reduces to $\varepsilon$ via CR, hence it is in $DQ_k$.
By definition of shuffle,  $w' \in  H_1\,   \shu\,  H_2$ has the form 
$x' \ff_j u' \ffc_j v'$ where $x'$ does not contain any closed parenthesis of $Dyck_{ff}$, 
the symbols $\ff_j , \ffc_j$ form the innermost pair in Dyck$_{ff}$ 
and  the factor $u'$ does not contain $\tfc_i$ or $\ft_i$. 
Therefore, CR 1 applies over and over until all symbols in $\Delta_{ff,k}$ are canceled.
Similarly, CR 2 cancels all parentheses in  $Dyck_{tt}$. Then $w'$ is reduced to a word $w''$ over AntiDyck symbols. Since $w'' \in H_3$, CR 3 and CR 4 
reduce it to $\varepsilon$.
Claim $DQ_k \subseteq H4$ : 
by Theor.~\ref{th:DQeqLDG}, there exists an LDG $(G,\lambda)$ such that $w\in W(\mathcal{G})$.
Consider the sequence $w_f$ of labels on  the  front row of $(G,\lambda)$ (e.g., in Fig.~\ref{fig:labeledDQ2graph} the subgraph with vertices $v'_1, \ldots, v'_n$). 
Clearly, word $w_f$ is over alphabet $\Delta_{ff,k} \cup \{\ft_1 ,\tfc_1, \ldots, \tfc_k,\}$ and belongs to language $H_1$. Similarly, the sequence of labels on the tail row 
(i.e., on vertices $v_1, \ldots, v_n$) is
a word $w_t$  in language $H_2$. Notice that $|w_f| + |w_t|=n$ and $w= w_f\, \shu \, w_t$, i.e.,  $w \in H_1\,   \shu \,  H_2$. 
It remains to prove that $w \in H_3$. This is immediate since the subgraph labeled with the projection of $w$ over 
$\Delta_{ft,k} \cup \Delta_{tf,k}$  contains all and only the queue edges, which do not cross and satisfy the definition of $H_3$. \qed
\begin{corollary}\label{cor:shuffleIntersectionForDA} 
For all languages $L$  in family \f, there exist context-free languages $L_1, L_2$, a language $L_Q$ recognized by a QRT queue automaton, and a rational transduction $\tau$ 
such that  $L= \tau\left((L_1\shu L_2)\, \cap\, L_3 \right).$
\end{corollary}

\paragraph{Linear deque graphs for generic deque languages.}\label{ssect:LDGforDA}
Since  each CDL word  is represented by an LDG (Theor.~\ref{th:DQeqLDG}) and from Theor.~\ref{theor:homomCharacteriz} each word of a  
language in \f is the homomorphic image of a  CDL word,  we can  equip   any deque language with a valuable graphical representation similar to CF syntax trees but more general.
\par
We loosely describe for Ex.~\ref{ex:dequelang}  how the  computation of $abbacccccc\in L_{pal}L_{sum}$  is converted into the LDG shown in Fig.~\ref{fig:DequeAutomataEx1}. 
For clarity, we denote $A$ with $A_{ff}$, $B$ with $B_{ff}$, etc. 
Scanning the first $a$, the DA pushes  
$A_{ff}$.  
The following moves are: scan $b$ and push $B_{ff}$, scan $a$ and pop $B_{ff}$.
The four preceding moves are represented by edges $(v'_1,v'_4)$ and $(v'_2,v'_3)$ which compose a so-called  ``rainbow'' pattern, essentially isomorphic to a syntax tree.
Now the  configuration is  $( q_1, ccccc, \varepsilon)$ and the next moves are: scan $c$ and write $C_{ft}D_{ft}$;  scan $c$, read $C$ and write $C_{ft}$;  scan $c$, read $D$ and write $C_{ft}D_{ft}$.
Now the configuration is $(p_3, ccc, C_{ft}C_{ft}D_{ft})$, the DA scans $ccc$ and empties the tape.
To be precise,  the two edges exiting $v'_5$ (and also $v'_7$) represent a non-simple DA transition. Such transitions where not present in $A_{CDL}$, and its LDG did not have vertices with non-hamiltonian degree 2 (already observed). 
But,  it would be immediate to normalize the DA in Fig.~\ref{fig:DequeAutomataEx1}, by introducing, between $v'_5$ and $v'_7$, a new vertex 
$v''_5$ with label $\varepsilon$ and  replacing  edge $(v'_5,v'_7)$ with edge $(v''_5,v'_7)$. The latter edge is associated 
with an $\varepsilon$-transition, and  corresponds to  an erasing value of the homomorphism of Theor.~\ref{theor:homomCharacteriz}. 

\par
We have seen that all \f\, languages have a syntax structure representable by toroidal embedding of planar  graphs having one hamiltonian path. In such LDG graphs it is possible to recognize subgraphs corresponding to  Dyck languages (\emph{ff} and \emph{tt}) and to AntiDyck languages (\emph{ft} and \emph{tf}); an illustration is the regular replicas language  defined in Prop.~\ref{prop-replicafamily}.
 Evidence for similar patterns occurs in the linear representation of  some  models for chemical  and biological molecular structures (e.g.,~\cite{barthel2015toroidal,Searls2002}).  
It is known that some of those patterns also occur in natural languages and are 
to some extent captured by existing grammar models, in particular by \emph{dependency grammars}.

\subsubsection{Conclusion.}
We have  shown that a deque used as  memory of an automaton, 
permits  computations beyond those possible with stacks and queues and  defines 
an interesting language family. 
Such computations are nicely represented by planar toroidal graphs, which facilitate intuitive  reasoning on deque languages.   
The basic closure and inclusion properties for deque automata have been established, but much remains to be done, in particular concerning  deterministic deque machines.
Comparisons with other existing models of grammars and automata featuring  stacks and queues remain for the future.  

\noindent{\bf Acknowledgements.} We thank the anonymous reviewers for their useful suggestions.

\bibliographystyle{abbrv}
\bibliography{automatabib}
\section*{Appendix: Proofs}
\noindent{\em Proof of Lemma~\ref{lm:normalForms}}
\begin{enumerate}
\item 
The proof in~\cite{DBLP:journals/tcs/CherubiniCCM91} that for every QRT nondeterministic queue automaton one can can effectively build an equivalent RT queue automaton is based on   encoding
a constant number of queue symbols into one, and then combining into one move (reading one input symbol)  the consecutive spontaneous steps. The same transformation  works 
for deque automata.
\item We construct a DA equivalent to $M$ with only one state $p$ and tape alphabet $\Gamma \cup \{\hat q \mid q \in Q\}$. 
$M'$ stores the current state $q$ of $M$ on the tape, into both the front and the tail symbols,  enforcing the following invariant: \\
 $(q, a u, \gamma_l\gamma\gamma_r)\to_M (q', u, \sigma\gamma\rho)
\iff (p, a u, \hat q\gamma_l\gamma\gamma_r \hat q )\to_{M'} (p, u, \hat q'\sigma\gamma\rho \hat q').
$ 
Notice that since $M$ accepts by empty tape, $M'$ must clear the tape from the two state-denoting symbols before accepting.

\item By definition, in a simple QRT DA, moves are restricted so that at each step there is only one deque operation on a single symbol, i.e., the automaton may 
perform either one read or one write operation on the deque, either at the front or at the tail (but not at both ends). 
It is obvious that, by adding suitable $\varepsilon$-transitions, a simple machine can simulate each move of a DA  that reads and/or writes on the deque a string of symbols.

\item 
The tape alphabet is first  duplicated in two sets, one for symbols written at the front and one for symbols written at the tail, also by suitably duplicating (for each nonterminal symbol) existing transitions that read from or write to the deque.  
A trivial application of nondeterminism (i.e., guessing at which end a symbol written on the deque will be subsequently read) allows to obtain the four  sets $\Gamma_{\ff}, \ldots$.
\item\label{it-zerotesting} 
 We show that by introducing a new tape symbol, $Z$, 
it is possible to test whether the deque is empty. 
Given a QRT deque automaton $M$ which has a spontaneous move for testing the deque for emptiness, we define an equivalent DA $M'$, with QRT delay $p+1$, without the ability of performing the deque emptiness, generalizing 
the construction given for queues in \cite{DBLP:journals/tcs/CherubiniCCM91}.  
$M'$ includes the original states of $M$ and all moves of $M$, except the move for testing emptiness; 
$M'$ has a new initial state, that  performs a spontaneous move writing $Z$ to the deque and then entering  the original initial state of $M$. 
For every state $q$ of $M$, add a new state $\tilde q$ such that if $(q,a,\gamma_l,\gamma_r,q',\gamma'_l,\gamma'_r)\in \delta$ is a transition of $M$, 
then $(\tilde q,a,\gamma_l,\gamma_r,q',\gamma'_l,\gamma'_r)$ is a
move of $M'$; also,
add the transitions $(q,\varepsilon,\varepsilon, Z, \tilde q, Z, \varepsilon)$ and $(q,\varepsilon,Z,  \varepsilon, \tilde q, \varepsilon, Z)$, to nondeterministically move $Z$ to the other tape end. 

To test whether the deque is empty while in a state $q$ and  then  moving to $q'$ whether it is empty and to $q''$ otherwise, we add a new state $\hat q$ and the transitions: 
\\
$(q,\varepsilon,\varepsilon, Z, \hat q, Z, \varepsilon)$, $(\hat q,\varepsilon, \varepsilon, Z, q',  \varepsilon,Z)$, $(\hat q,\varepsilon,A,  \varepsilon, q'', \varepsilon,A)$ for every $A \in\Gamma -\{Z\}$. 
\\
At last, define a new DA $M''$, accepting by final state,  which is identical to $M'$
but with the ability of testing for emptiness: $M''$ has however only one final (new) state $p$ and  
for every state of $M''$ corresponding to a final state of $M'$ 
$M''$ has an $\varepsilon$-transition to $p$ deleting $Z$, i.e., that can be taken only if the deque is empty. 
\end{enumerate}
\qed

\noindent{\em Proof of Proposition~\ref{prop-closureHomRevhom}.}
It is enough to show closure under union, intersection with regular languages, 
nonerasing homomorphism,  reverse homomorphism, 
and 
Kleene closure, and non closure under erasing homomorphism.
\begin{enumerate}
 \item Closures under union and under intersection with regular languages are obvious. 

\item   Closure under nonerasing homomorphism: let $L$ be a language in \f. 
If $h$ is the homomorphism mapping a letter $a \in \Sigma$ to a word $w_a$ over a finite alphabet $\Delta$, consider
the SDA accepting $L$. If $|w_a|= 1$, just replace $a$ with $w_a$ in every transition of the SDA. If $|w_a|=m>1$, then 
for each state $q$ introduce $m$ new states $q_{a_1}, \dots, q_{a_{m-1}}$, so that each transition of the form $(q,a,q', X)$ 
(where $X$ is one of $\overleftarrow{A_{ff}}$, etc.) is replaced by the transitions 
\[(q,w_a(1),q_{a_1},X), (q_{a_1},w_a(2),q_{a_2},\varepsilon) \dots,  (q_{a_{m-1}},w_a(m),q',\varepsilon).\]



\item Closure under reverse homomorphism. We do not have to write the proof in drtail, since it is identical to the traditional proof of the same result for pushdown automata. Let $g: \Sigma \to \Delta^*$ be a (possibly erasing) homomorphism, let $L\in \text{\f} $ be a language over $\Delta$ and let $M$ be an SDA recognizing $L$. 
Let $p=\max{\{|g(a)| : a \in \Sigma\}}$ and build a DA $M'$ over alphabet $\Sigma$, based on the idea to use the finite state control as a buffer of length $p$ to
store $p$ symbols of alphabet $\Delta$. When reading an input letter $a \in \Sigma$, $M'$ simulates $M$ on the homomorphic image $g(a)$, which is stored in the buffer, using up to $p$ $\varepsilon$-transitions.
 If $M$ accepts its input, then also $M'$ accepts, i.e., $M'$ accepts $g^{-1}(L)$.

\item 
Since testing for empty deque can be obtained (as explained in Part~\ref{it-zerotesting} of the proof of Lemma~\ref{lm:normalForms}) by adding a new special tape symbol, the closure under Kleene star is obvious. 
In fact, let $M$ be a QRT deque automaton, with the ability of testing the deque for emptiness, recognizing a language $L$;  assume, without loss of generality,  that there is no $\varepsilon$-transition from the initial state. Define a  DA $M'$ with the same states and transitions of $M$; 
its set of final states is the set of final states of $M$ plus the initial state of $M$, and such that each final but not initial state of $M'$ has also an outgoing $\varepsilon$-transition back to the the initial state, 
testing whether the deque is empty. 

\item To prove nonclosure under erasing homomorphism, let $Q^*_2$ be the AntiDyck language  on an alphabet containin two brackets; $Q^*_2$ is obviously in \f, and is the generator of the trio (rational cone) of all recursively enumerable languages \cite{DBLP:journals/tcs/VauquelinF80}. 
The family \f is closed under inverse homomorphism and intersection with regular languages, therefore, if \f were closed also under erasing homomorphism, then it should include all recursively enumerable languages;  
\f is included in the family of 4-tape RT Turing machines~\cite{jacm/LeongSeiferas81}. 
\end{enumerate}
\qed

\noindent{\em Proof of Proposition~\ref{prop-replicafamily}.}
The language $L$ given in the proof hint can be defined by DA
\[
M= (\Sigma\cup\{D,R\},\Gamma,\{q_0,q_T,q_F,q'_T,q'_F,q_\textit{fin}\}, q_0, \delta,\{q_\textit{fin}\})
\]
where $\Gamma= \Sigma \cup Z$, with $Z\not\in\Sigma$, and $\delta$ is defined next.  We assume for simplicity 
$\Sigma=\{a,b\}$. 
\par\noindent
In the initial state $q_0$, given a word $uD\dots$ or $uR\dots$, $M$ stores $u$ in the deque, so that when reading from the front  it reads $u$, when reading from the tail it reads $u^R$.
Upon scanning $D$ or $R$, $M$ changes its state to the state $q_F$ or, respectively, to the state $q_T$ and writes $Z$ to the tail or, respectively, to the front. 
A first batch of transitions in $\delta$ are thus the following:
\[
(q_0,a,\varepsilon,\varepsilon,q_0,\varepsilon,a),(q_0,b,\varepsilon,\varepsilon,q_0,\varepsilon,b),(q_0,D,\varepsilon,\varepsilon,q_F,\varepsilon,Z),(q_0,R,\varepsilon,\varepsilon,q_T,Z,\varepsilon)
\]
In state $q_F$, $M$ compares the current input symbol $a$ with the symbol at the front: 
if they are different it crashes, else it writes $a$ to the tail and continues reading the next symbol, 
until $D,R$ is found and $Z$ is read from the front -- at that point, the deque content is still $u$; 
$M$ writes $Z$ to the front or to the tail and changes to state $Q_T$ or $q_F$ respectively. 
The behavior in state $q_T$ is symmetrical, but by reading from the tail it reads $u^R$. The corresponding transitions in $\delta$ are the following:
\[
 \begin{array}{llll}
(q_F,a,a,\varepsilon,q_F,\varepsilon,a),(q_F,b,b,\varepsilon,q_F,\varepsilon,b),(q_F,D,Z,\varepsilon,q_F,\varepsilon,Z),(q_F,R,Z,\varepsilon,q_T,Z,\varepsilon) \\
(q_T,a,\varepsilon,a,q_T,a,\varepsilon),(q_T,b,\varepsilon,b,q_T,b,\varepsilon),(q_T,D,\varepsilon,Z,q_F,\varepsilon,Z),(q_T,R,\varepsilon,Z,q_T,Z,\varepsilon) \\
\end{array}
\]
Machine $M$ nondeterministically guesses when the current factor is the last one: 
it stops storing the replica $u$ on the tape (using the two states $q'_T, q'_F$ that only compare the tape content with the input), so that at the end of the word the deque 
only contains $Z$, which can be deleted upon entering 
the final state $q_\textit{fin}$. The transitions in $\delta$ for this part of the computation are:
\[
 \begin{array}{llll}
(q_0,D,\varepsilon,\varepsilon,q'_F,\varepsilon,Z),(q_0,R,\varepsilon,\varepsilon,q'_T,Z,\varepsilon)
(q_F,D,Z,\varepsilon,q'_F,\varepsilon,Z),(q_F,R,Z,\varepsilon,q'_T,Z,\varepsilon)\\
(q_T,D,\varepsilon,Z,q'_F,\varepsilon,Z),(q_T,R,\varepsilon,Z,q'_T,Z,\varepsilon)\\
(q'_F,a,a,\varepsilon,q'_F,\varepsilon,\varepsilon),(q'_F,b,b,\varepsilon,q'_F,\varepsilon,\varepsilon),(q'_F,Z,\varepsilon,q_\textit{fin},\varepsilon,\varepsilon)\\
(q'_T,a,\varepsilon,a,q_T,\varepsilon,\varepsilon),(q'_T,b,\varepsilon,b,q'_T,\varepsilon,\varepsilon),(q'_T,\varepsilon,Z,q_\textit{fin},\varepsilon,\varepsilon)
\end{array}
\]
An example of computation is :
\[
\begin{array}{l}
(q_0,abbDabbDabbRbbaRbba,\varepsilon)\stackrel{abb}{\to} 
(q_0,DabbDabbRbbaRbba,abb)\stackrel{D}{\to}\\
(q_F,abbDabbRbbaRbba,abbZ)\stackrel{abb}{\to}(q_F,DabbRbbaRbba,Zabb) \stackrel{D}{\to}\\ 
(q_F,abbRbbaRbba,abbZ)\stackrel{abb}{\to}(q_F,RbbaRbba,Zabb)\stackrel{R}{\to} \\
(q_T,bbaRbba,Zabb)\stackrel{bba}{\to}(q_T,Rbba,abbZ)\stackrel{R}{\to} \\
(q'_T,bba,Zabb)\stackrel{bba}{\to}(q'_T,\varepsilon,Z)
\stackrel{\varepsilon}{\to}(q_\textit{fin},\varepsilon,\varepsilon).
\end{array}
\]
%

\noindent{\em Proof of Th.~\ref{th:DQeqLDG}.}
We first define more precisely the deque graph machine DGM defined in~\cite{DBLP:books/daglib/0035667} as a recognizer of deque graphs.
Given a graph $G$ with a Hamiltonian path $p$, a DGM walks on  $p$, reading at every step a vertex and its associated edge -- as already noticed, here we consider 
graphs with exactly  one edge per vertex
(excluding  the edge of the Hamiltonian path). 
If the edge is outgoing, then it is written 
to the front or to the tail of the deque, according to the type of the edge;
if the edge is incoming, then it is read from the deque, from the front or from the tail according to the type of the edge -- if not present in the deque at the correct position, then the DGM halts without accepting. 
\par
Since we want to consider a labeled deque graph LDG rather than a deque graph (without labels), we also need a finite alphabet, which is absent in the original definition~\cite{DBLP:books/daglib/0035667}, 
to label each node of the graph -- the alphabet we consider here is precisely $\Delta_k$.
In this way, we  define a new version of the DGM, called a {\em $\Delta_k$-labeled DGM}, or $\Delta_k$-DGM for short, that recognizes the family of graphs in $\mathcal{G}_k$.

\par
In addition to the behavior of a DGM, a $\Delta_k$-DGM must  make sure that, for each edge, the labels of both vertices are correct. 
Thus, when writing an edge $\langle \alpha, \beta\rangle$ to the deque (e.g., by writing the two vertices in the deque), 
it must also store the label $\lambda(\alpha)$ of vertex $\alpha$.
then, upon reading the edge $\langle \alpha, \beta\rangle$ from  deque, it must  verify  not only that the current vertex is equal to $\beta$, 
but also that its label $\lambda(\beta)$ is the closing version of $\lambda(\alpha)$; if not, the machine  halts without accepting. 
It is clear that, given $k\ge 1$, a $\Delta_k$-DGM recognizes the family $\mathcal{G}_k$.
Such machine can also be seen as accepting words, i.e., the words of the form $W(G,\lambda)$ for some graph  $G \in \mathcal{G}_k$.
\par
The proof considers separately the two inclusions.
\begin{description}
 \item[$W(G,\lambda) \subseteq DQ_k$:]
 
Let $(G,\lambda)\in \mathcal{G}_k$ be a LDG, for some $k\ge 1$, and let $w = W(G,\lambda)$. We prove that $w \in DQ_k$.

Given a prefix $v$ of the word $W(G,\lambda)$ on the hamiltonian path, let $\rho(v)$ be the content of the deque of $\Delta_k$-DGM after reading the first $|v|$ vertices of 
$(G,\lambda)$. Similarly, let $\gamma(v)$ be the  content of the deque of $A_{CDL}$ after reading a prefix $v$ of its input.

We show by induction on $|v|\ge 0$, with $v$ a prefix of  $w$, that 
the content of the deques is such that $\gamma(v) = \Lambda(\rho(v))$,
where $\Lambda: E \to  \Gamma$ is the homomorphism defined as the ``upper case'' version of $\lambda$, i.e., 
if $\lambda(e) =\ffc_j$ then $\Lambda(e) = \FF_j$, etc. 
The thesis that $w \in DQ_k$ is then an immediate consequence (since both machines accept with empty deque).
\par\noindent
The base case is $|v|=0$, i.e., when both machines are in the initial configuration: both deques are just empty.
Assume now that the induction hypothesis holds for a prefix $v$ of $w$, with $0\le|v|<|w|$.
When reading the next vertex $\alpha$ of $(G,\lambda)$ along the hamiltonian path, corresponding to an edge $e$ 
we have two cases.

\begin{enumerate}

\item If $e$ is an outgoing edge (i.e., it has the form $\langle \alpha,\beta\rangle$), 
then 
 $\lambda(\alpha)$ is an $\overrightarrow{a_j}\in\overrightarrow{\Delta}_k$;

$\Delta_k$-DGM writes the edge $e$ 
to the deque, 
either at the tail (if 
$\overrightarrow{a_j}$ is  $\tto_j$ or $\tf_j$) 
or at the front (if 
$\overrightarrow{a_j}$ is  $\ff_j$ or $\ft_j$); 
by definition, $A_{CDL}$ writes $\Lambda(\overrightarrow{a_j})$ at the same end -- front or tail, depending on $\overrightarrow{a_j}$.  
Clearly, if writing to the front, then $\gamma(v\overrightarrow{a_j})= \overrightarrow{a_j}\gamma(v) = \Lambda(e) \Lambda(\rho(v)) 
= \Lambda(\rho(v\overrightarrow{a_j})$; 
the case when writing to the tail is symmetrical: $\gamma(v\overrightarrow{a_j})=\gamma(v)\overrightarrow{a_j} = 
\Lambda(\rho(v))\Lambda(e)=\Lambda(\rho(v\overrightarrow{a_j})$. 
 \item 
If $e$ is an incoming  edge (i.e., it has the form $\langle \beta,\alpha\rangle$), then 
  $\lambda(\alpha)$ is an  $\overleftarrow{a_j} \in\overleftarrow{\Delta}_k$.

The $\Delta_k$-DGM must read the edge $e$ from the tail or from the front of the deque (depending on the vertex label).
Hence,  $\rho(v)$ has either the form $e_1 \dots e_he$ or $e e_1 \dots e_h$ (for some $h\ge 0$ and edges $e_1, \dots, e_h$); 
therefore reading edge $e$ only $e_1 \dots e_h$ is left  in the deque, i.e., $\rho(v\overleftarrow{a_j})= e_1 \dots e_h$.
By induction hypothesis, $\gamma(v) = \Lambda(\rho(v))=\Lambda(e_1 \dots e_h e)$, which is equal 
to $\Lambda(e_1) \dots \Lambda(e_h) \Lambda(e)= \Lambda(e_1) \dots \Lambda(e_h)\Lambda(\overleftarrow{a_j})$ 
(symmetrically 
$\gamma(v) = \Lambda(\overleftarrow{a_j})\Lambda(e_1) \dots \Lambda(e_h)$ holds if reading from the front). 
Then, $A_{CDL}$ with deque content  $\gamma(v)$ can make a move reading $\overleftarrow{a_j}$ from the input and consuming 
$\Lambda(\overleftarrow{a_j})$ from the deque: the new deque content $\gamma(v\overleftarrow{a_j})$ is 
thus $\Lambda(e_1) \dots \Lambda(e_h)= \Lambda(\rho(v\overleftarrow{a_j}))$.
\end{enumerate}

\item[$DQ_k\subseteq W(G,\lambda)$:] Let $w \in DQ_k$, for some $k\ge 1$. 
Consider a run of $A_{CDL}$ over $w$ and, as in the proof of the previous part, let $\gamma(v)$ be the content of the deque  after reading a  prefix $v$ of $w$.
Given a graph $(G,\lambda)$, let $\rho(v)$ be the content of the deque of $\Delta_k$-DGM 
after reading the first $|v|$ vertices along the hamiltonian path, labeled with word $v$.
\par\noindent
We prove by induction on $n>0$, that for every $w \in DQ_k$, with $|w|\le n$, 
there exists a LDG $(G,\lambda)$, with $|w|$ vertices, such that:  

\begin{enumerate}
\item $W(G,\lambda)=w$;
\item For every prefix $v$ of $w$, $\gamma(v) = \Lambda(\rho(v))$. 
\end{enumerate}

%
The base case $n=0$ is trivial. Assume the induction hypothesis holds for $n\ge 0$. 
Consider a word $w \in DQ_k$ such that $|w|=n+2$ (the length of words in $DQ_k$ is always even).  
Hence, $w$ has the form $x\overrightarrow{a_j} y\overleftarrow{a_j} z$, with $xyz \in DQ_k$ and
$\overrightarrow{a_j},  \overleftarrow{a_j}\in \Delta_k$. 
The induction hypothesis applies to $xyz$: let $(G,\lambda)$ be an LDG such that $W(G,\lambda)=xyz$.

Consider a labeled graph $(G',\lambda)$ obtained from $(G,\lambda)$, by adding two new vertices $\alpha, \beta$ (as explained next) 
and a new edge $e= \langle \alpha, \beta\rangle$, and by
extending $\lambda$ by letting $\lambda(\alpha)= \overrightarrow{a_j}$, $\lambda(\beta)= \overleftarrow{a_j}$.

Vertices $\alpha$ and $\beta$  are inserted  along the hamiltonian path: $\alpha$ immediately after the first $|x|$ vertices of $G$, and 
$\beta$ immediately after the first $|xy|$ vertices of $G$. It is obvious that 
$W(G',\lambda)=x\overrightarrow{a_j}y\overleftarrow{a_j}z=w$. We need to show that this labeled graph is still a LDG, 
i.e., it is accepted by a $\Delta_K$-DGM.

Since by induction hypothesis $\gamma(x) = \Lambda(\rho(x))$, 
it is also true that $\gamma(x\overrightarrow{a_j}) = \Lambda(\rho(x\overrightarrow{a_j}))$, because, immediately after 
reading the vertex $\alpha$, 
 machine $\Delta_k$-DGM writes the edge $e$ to the deque, in the same position where 
$A_{CDL}$ writes $\Lambda(\overrightarrow{a_j})=\Lambda(e)$. 
After reading also $y$ from the input, the deque of $A_{CDL}$ 
is $\gamma(x\overrightarrow{a_j} y)$, which is equal to  
$\Lambda(\rho(x\overrightarrow{a_j}y))$
(since by induction hypothesis
also $\gamma(xy) = \Lambda(\rho(xy))= \Lambda(\rho(x))\Lambda(\rho(y))$). 
The deque of $A_{CDL}$ must have, when reading $\overleftarrow{a_j}$, the tape symbol $\Lambda(\overleftarrow{a_j})$ at the front or at the tail (depending on the letter), thus 
the deque of $\Delta_k$-DGM  must have $e$ at the same position. Therefore, the latter machine may read vertex $\beta$, consuming the edge $e$ from the deque.
After this move, the deque of the two machines is, respectively, $\gamma(x\overrightarrow{a_j} y\overleftarrow{a_j})=\gamma(xy)$ and 
$\rho(x\overrightarrow{a_j} y \overleftarrow{a_j}) = \rho(xy)$, hence $\Delta_k$-DGM accepts also $(G',\lambda)$.
\end{description}

Thus, $W(\mathcal{G}) = DQ_k$.
 
\qed

\noindent{\em Proof of Th.~\ref{theorCancellationRules}.}
Part 1. 
Let $w\subseteq {\Delta_k}^*$ and $w\not\in DQ_k$. We show that the word $w'$, resulting from application of a CR to $w$, is not in $DQ_k$. 
The thesis then follows by a simple induction. It suffices to examine CR 1 and 3 since the others are symmetrical. 
 

If CR 1. applies, then $w=u \,\ff_i \, x\, \ffc_i \, v $ and $w'= u xv$ as in the definition of CR 1. 
Assume that, reading $w$, $A_{CDL_k}$ may scan prefix $u$ and still reach a valid configuration;  otherwise if every computation of $A_{CDL_k}$ blocks before reading $\ff_i$, 
then it also blocks in the same position reading $w'=uxv$.

Clearly, the $|u|+1$-th   move  writes $\FF_i$ on the front, while each move reading the segment $x$ may only operate at the tail,  performing one  of the actions:   
store some $\TF_j$ or $\TT_j$ in the tail, read some $\FT_j$ or $\TT_j$ from the tail. Since those moves do not operate at the front, 
the successive move scanning $\ffc_i$  can still read $\FF_j$ at the front, deleting from the deque.
Therefore, after reading $u\,\ff_i \, x\, \ffc_i$ or $u x $ machine $A_{CDL_k}$ may reach the same configurations. 

%

If CR 3. applies, then $w=\ft_i\,x \,\ftc_i \, v$ and $w'= xv$ as in the definition of CR 3. 
The first move of $A_{CDL_k}$ stores $\FT_i$ at the front. Since $x$ has only symbols of the form $\ft_j$, 
that correspond to a write action of $\FT_j$ at the tail, 
we can  assume that, while reading $w$, $A_{CDL_k}$ may scan prefix $x$ and still reach a valid configuration, else every computation would block 
also reading $w'$. It is obvious that after reading $x$, $\FT_i$ is still the top of tail, so that the move scanning $\ftc_i$, which reads at tail,  consumes  $\FT_i$.
The configurations that may be reached scanning $\ft_i\,x \,\ftc_i$ or $x$ are the same, i.e., with the same deque content and $v$ still to be scanned. 
Therefore, if $A_{CDL_k}$ does not accept $w$, it can neither accept $w'$. 


Part 2. Let $w\in DQ_k$, then $w$ reduces to $\varepsilon$ via CR.
From Th.~\ref{th:DQeqLDG} we have that $w$ is the labeled Hamiltonian path of an LDG. It is immediate to check  that each CR exactly cancels an edge (actually the leftmost edge) of such graph. 
\qed

\end{document}